\documentclass[twoside,twocolumn,9pt]{article}
\usepackage{extsizes}
\usepackage[super,sort&compress,comma]{natbib} 
\usepackage[version=3]{mhchem}
\usepackage[left=1.5cm, right=1.5cm, top=1.785cm, bottom=2.0cm]{geometry}
\usepackage{balance}
\usepackage{mathptmx}
\usepackage{sectsty}
\usepackage{graphicx} 
\usepackage{lastpage}
\usepackage[format=plain,justification=justified,singlelinecheck=false,font={stretch=1.125,small,sf},labelfont=bf,labelsep=space]{caption}
\usepackage{float}
\usepackage{fancyhdr}
\usepackage{fnpos}
\usepackage[english]{babel}
\addto{\captionsenglish}{%
  
}
\usepackage{array}
\usepackage{droidsans}
\usepackage{charter}
\usepackage[T1]{fontenc}
\usepackage[usenames,dvipsnames]{xcolor}
\usepackage{setspace}
\usepackage[compact]{titlesec}
\usepackage{hyperref}
\usepackage{epstopdf}%This line makes .eps figures into .pdf - please comment out if not required.

\definecolor{cream}{RGB}{222,217,201}
\usepackage{amsmath,amssymb,graphicx}
\usepackage{xcolor}  %for colored text: \textcolor{red}{text to be colored}
\usepackage{soul} % for striking through text: \st{text to be struck throw }
\usepackage{upgreek}

\begin{document}

\pagestyle{fancy}
\thispagestyle{plain}
\fancypagestyle{plain}{
%%%HEADER%%%
\renewcommand{\headrulewidth}{0pt}
}
%%%END OF HEADER%%%

%%%PAGE SETUP - Please do not change any commands within this section%%%
\makeFNbottom
\makeatletter
\renewcommand\LARGE{\@setfontsize\LARGE{15pt}{17}}
\renewcommand\Large{\@setfontsize\Large{12pt}{14}}
\renewcommand\large{\@setfontsize\large{10pt}{12}}
\renewcommand\footnotesize{\@setfontsize\footnotesize{7pt}{10}}
\makeatother

\renewcommand{\thefootnote}{\fnsymbol{footnote}}
\renewcommand\footnoterule{\vspace*{1pt}% 
\color{cream}\hrule width 3.5in height 0.4pt \color{black}\vspace*{5pt}} 
\setcounter{secnumdepth}{5}

\makeatletter 
\renewcommand\@biblabel[1]{#1}            
\renewcommand\@makefntext[1]% 
{\noindent\makebox[0pt][r]{\@thefnmark\,}#1}
\makeatother 
\renewcommand{\figurename}{\small{Fig.}~}
\sectionfont{\sffamily\Large}
\subsectionfont{\normalsize}
\subsubsectionfont{\bf}
\setstretch{1.125} %In particular, please do not alter this line.
\setlength{\skip\footins}{0.8cm}
\setlength{\footnotesep}{0.25cm}
\setlength{\jot}{10pt}
\titlespacing*{\section}{0pt}{4pt}{4pt}
\titlespacing*{\subsection}{0pt}{15pt}{1pt}
%%%END OF PAGE SETUP%%%

%%%FOOTER%%%
\fancyfoot{}
\fancyfoot[LO,RE]{\vspace{-7.1pt}\includegraphics[height=9pt]{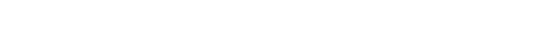}}
\fancyfoot[CO]{\vspace{-7.1pt}\hspace{13.2cm}\includegraphics{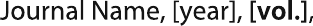}}
\fancyfoot[CE]{\vspace{-7.2pt}\hspace{-14.2cm}\includegraphics{head_foot/RF}}
\fancyfoot[RO]{\footnotesize{\sffamily{1--\pageref{LastPage} ~\textbar  \hspace{2pt}\thepage}}}
\fancyfoot[LE]{\footnotesize{\sffamily{\thepage~\textbar\hspace{3.45cm} 1--\pageref{LastPage}}}}
\fancyhead{}
\renewcommand{\headrulewidth}{0pt} 
\renewcommand{\footrulewidth}{0pt}
\setlength{\arrayrulewidth}{1pt}
\setlength{\columnsep}{6.5mm}
\setlength\bibsep{1pt}
%%%END OF FOOTER%%%

%%%FIGURE SETUP - please do not change any commands within this section%%%
\makeatletter 
\newlength{\figrulesep} 
\setlength{\figrulesep}{0.5\textfloatsep} 

\newcommand{\topfigrule}{\vspace*{-1pt}% 
\noindent{\color{cream}\rule[-\figrulesep]{\columnwidth}{1.5pt}} }

\newcommand{\botfigrule}{\vspace*{-2pt}% 
\noindent{\color{cream}\rule[\figrulesep]{\columnwidth}{1.5pt}} }

\newcommand{\dblfigrule}{\vspace*{-1pt}% 
\noindent{\color{cream}\rule[-\figrulesep]{\textwidth}{1.5pt}} }

\makeatother
%%%END OF FIGURE SETUP%%%

%%%TITLE, AUTHORS AND ABSTRACT%%%
\twocolumn[
  \begin{@twocolumnfalse}
{\includegraphics[height=30pt]{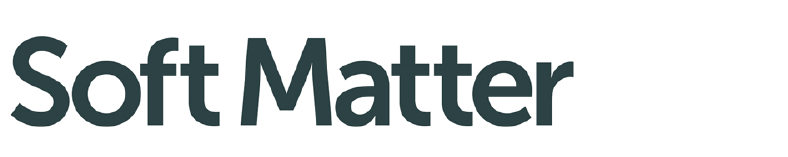}\hfill\raisebox{0pt}[0pt][0pt]{\includegraphics[height=55pt]{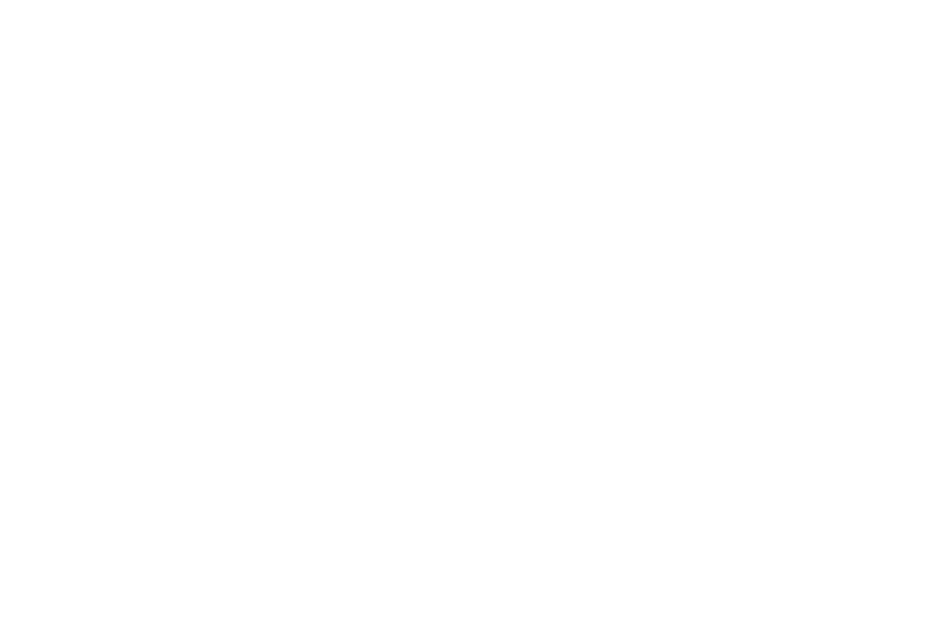}}\\[1ex]
\includegraphics[width=18.5cm]{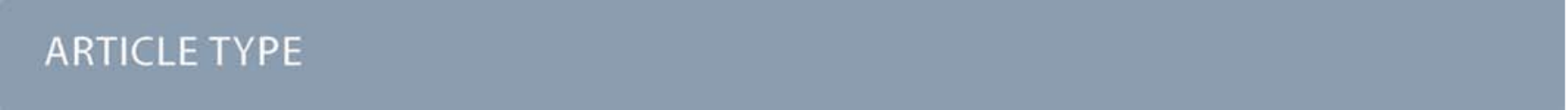}}\par
\vspace{1em}
\sffamily
\begin{tabular}{m{4.5cm} p{13.5cm} }

\includegraphics{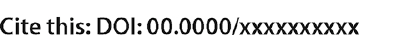} & \noindent\LARGE{\textbf{Effect of colloidal weight at different planar interfaces}} \\%Article title goes 

\vspace{0.3cm} & \vspace{0.3cm} \\

% & \noindent\large{Full Name,$^{\ast}$\textit{$^{a}$} Full Name,\textit{$^{b\ddag}$} and Full Name\textit{$^{a}$}} \\%Author names go here instead of "Full name", etc.
& \noindent\large{Mehdi Shafiei Aporvari,$^{\ast}$\textit{$^{abc}$} Agnese Callegari,\textit{$^{d}$} and Emine Ulku Saritas\textit{$^{bce}$}} \\%Author names go here instead of "Full name", etc.

\includegraphics{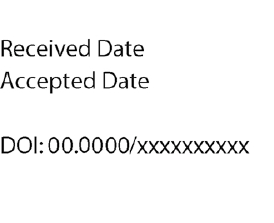} & \noindent\normalsize{In many physical and biological systems, particles and microorganisms move in the proximity of an interface. Understanding the dynamics of a particle suspended close to an interface is not only important conceptually but is crucial for practical applications ranging from the treatment of waste waters to industrial applications of self-assemblies.
In this work, we experimentally investigate the effects of colloidal weight on its dynamics while moving %\st{in close contact with} 
in the close proximity of a variety of liquid-solid and liquid-air interfaces.
Using an upward magnetic force, we change the effective weight of a superparamagnetic colloid. 
%and study its diffusivity in a proximity of interfaces. 
At water-glass interfaces, we observe the expected decrease of the diffusion coefficients with increasing effective weight.
At liquid-air interfaces, while for a pure water-air interface there is a negligible dependency between the diffusivity and the effective weight, we find that in semi-dilute polymer solution  at the solution-air interface  the diffusivity of the particle shows similar behavior as those at liquid-solid interfaces.
 We implement a Brownian dynamics simulation to support our results and show how the interplay between  hydrodynamic interactions and electrostatic interface repulsion explains the experimental results. } \\%The abstrast goes here instead of the text "The abstract should be..."

\end{tabular}

 \end{@twocolumnfalse} \vspace{0.6cm}

  ]
%%%END OF TITLE, AUTHORS AND ABSTRACT%%%

%%%FONT SETUP - please do not change any commands within this section
\renewcommand*\rmdefault{bch}\normalfont\upshape
\rmfamily
\section*{}
\vspace{-1cm}

%%%FOOTNOTES%%%

\footnotetext{\textit{$^{a}$~Department of Physics \& Biophysics, University of San Diego, San Diego, CA 92110, USA. E-mail: maporvari@sandiego.edu}}

\footnotetext{\textit{$^{b}$~UNAM -- National Nanotechnology Research Center, Bilkent University, Ankara 06800, Turkey.}}
\footnotetext{\textit{$^{c}$~National Magnetic Resonance Research Center (UMRAM), Bilkent University, Ankara 06800, Turkey.}}
\footnotetext{\textit{$^{d}$~Department of Physics, University of Gothenburg, SE-41296 Gothenburg, Sweden.}}
\footnotetext{\textit{$^{e}$~Department of Electrical and Electronics Engineering, Bilkent University, Ankara, Turkey.}}

%Please use \dag to cite the ESI in the main text of the article.
%If you article does not have ESI please remove the the \dag symbol from the title and the footnotetext below.
%\footnotetext{\dag~Electronic Supplementary Information (ESI) available: [details of any supplementary information available should be included here]. See DOI: 10.1039/cXsm00000x/}
%additional addresses can be cited as above using the lower-case letters, c, d, e... If all authors are from the same address, no letter is required

%\footnotetext{\ddag~Additional footnotes to the title and authors can be included \textit{e.g.}\ `Present address:' or `These authors contributed equally to this work' as above using the symbols: \ddag, \textsection, and \P. Please place the appropriate symbol next to the author's name and include a \texttt{\textbackslash footnotetext} entry in the the correct place in the list.}

%%%END OF FOOTNOTES%%%

%%%MAIN TEXT%%%%
%**************************************************************************************************
%**************************************************************************************************
%**************************************************************************************************

\section{Introduction}
 The motion of a spherical particle parallel to a plane wall, which was   first treated by Faxen \citep{faxen1922widerstand},  has been studied extensively in many theoretical and experimental works
 \citep{huang2015effect, lin2000direct, cichocki2004motion, lee1979motion, ha2013direct, villa2020motion}.
Close to an interface, the interplay between the hydrodynamics, electrostatics and gravitational forces leads to completely different dynamics  from those in the bulk. 
Upon approaching a planar interface, the parallel ($D_{||}$) and perpendicular ($D_{\perp}$)  diffusion coefficients of a spherical particle change dissimilarly compared to the bulk value $(D_0)$.
In particular, the diffusivity in the direction parallel to the planar interface  could either increase or decrease depending on the surface boundary conditions \citep{happel2012low}. At solid-liquid interfaces parallel diffusivity reduces as the particle approaches to the interface, which is referred to as  {\em hindered diffusion} \citep{bevan2000hindered, sharma2010high, kazoe2011measurements}. In the case of liquid-air interface with perfect slip boundary condition, the parallel diffusion increases to values even  bigger than the corresponding bulk diffusion coefficients \citep{nguyen2004exact}.

Apart from the asymmetric change in diffusivity, which stems from  hydrodynamics interactions, external forces such as gravity also play an important role in the dynamics of a particle moving close to an interface.
At low Reynolds number, the dynamics of a colloidal particle suspended in the bulk of a liquid is not affected by inertia and the effect of its weight on the dynamics is negligible on the typical time scales of  its Brownian motion. On the contrary, when the particle is close to an interface, its weight  determines its equilibrium distance from the interface. In such a case, differently from what happens in the bulk, the weight directly affects the dynamics of the particle:  in fact, the mobility of the colloid depends on the
frictional hydrodynamic forces, which in turn depend on the particle distance from the interface,  determined by its weight.
The dynamics of the colloidal particle is also strongly influenced by the boundary condition at the interface itself, which depends on the specific kind of interface (solid-liquid, liquid-air) and on other factors like the presence of impurities. It is, in fact, well known that impurities in the liquid surface can change the perfect slip boundary condition to a partial slip boundary condition \citep{maali2017viscoelastic, manor2008dynamic, manor2008hydrodynamic}.

Optical tweezers and interferometry are standard  techniques to study the dynamics of a colloidal particle near an interface experimentally \cite{sharma2010high, kazoe2011measurements, bevan2000hindered}. 
For example, using optical tweezers, it was shown that near a water-air interface the particle drag coefficients reduces with respect to the bulk value, while near a water-solid interface the drag coefficient is enhanced \cite{wang2009hydrodynamic}. 
The interaction energy between a  sedimented microparticle and water-air interface has been studied experimentally using reflecting wave microscopy \citep{villa2020multistable}.

In this work, we use an upward magnetic field to control the effective weight of a superparamagnetic colloidal particle in the proximity of a liquid-solid or liquid-air interface.
 We implement a high-frame-rate video microscopy to record the trajectories of isolated particles at different interfaces. We show experimentally that the diffusivity of a colloidal particle in close contact with a liquid-glass interface changes considerably by reducing the effective weight. 
Also in agreement with the theoretical model, our experimental results show that, for a frictionless interface like water-air interfaces,  the diffusivity in the parallel direction does not noticeably change by changing the effective weight.  
Interestingly, we show that, at the interface between an aqueous semi-dilute polymer solution and air, the motion of the particles resembles more to that of particles at liquid-solid interfaces than at liquid-air interfaces. 
Our experimental results are in agreement with the results of Brownian dynamics simulations
 of a particle in the gradient of diffusion under the effect of weight, buoyancy, and electrostatic colloidal interaction.

\begin{figure}
\includegraphics[width=1.0\columnwidth]{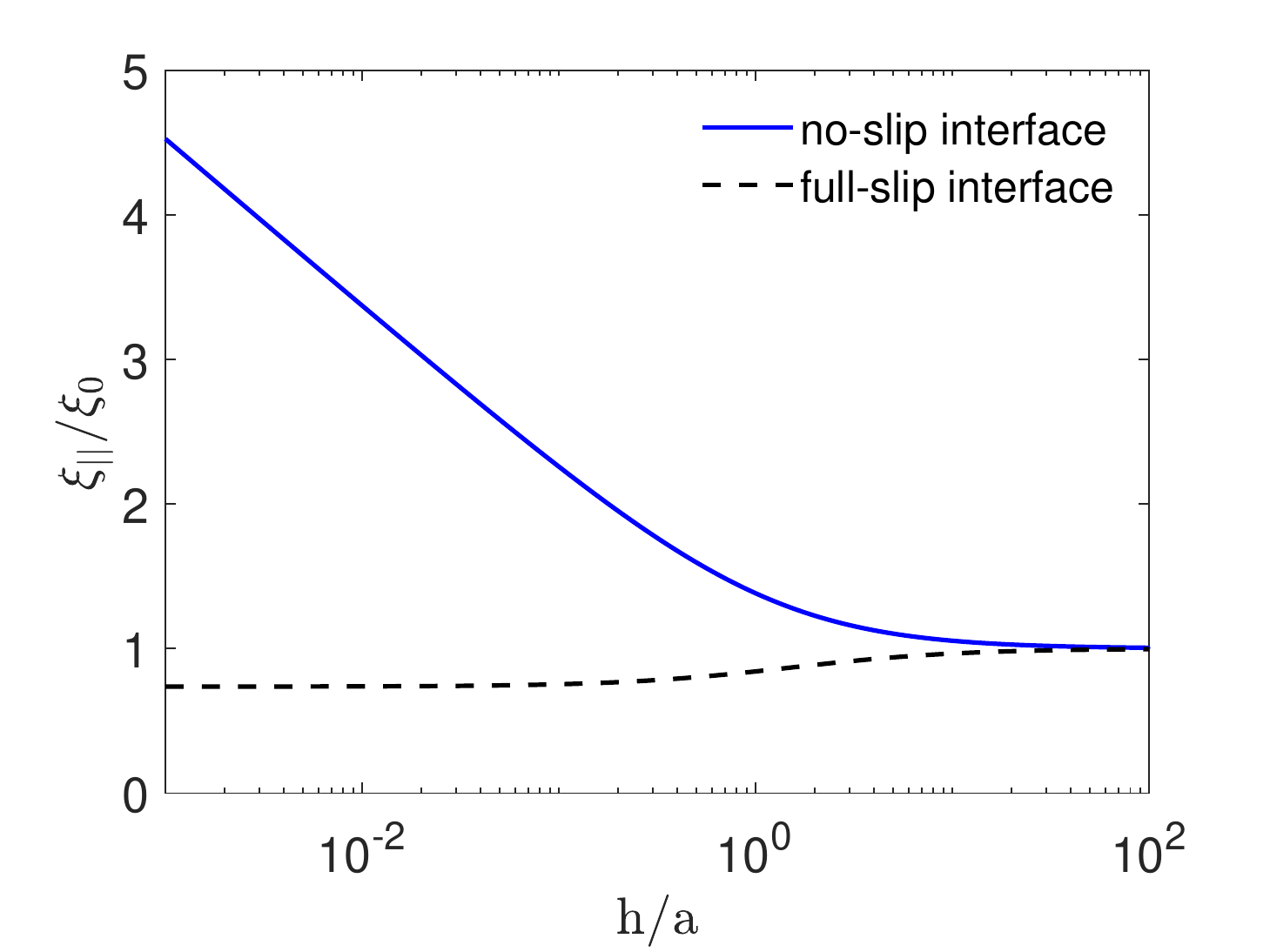}
\caption{Drag coefficients in the parallel direction ($\xi_{||}$) close to a planar interface for no-slip (solid line, Eq. (1)) and full-slip (dashed line, Eq. (2)) boundary conditions. The drag coefficient $\xi_{||}$ is expressed in units of the drag coefficient in the bulk $\xi_{0}$ as a function of $h/a$, where $a$ is the particle radius and $h$ is the gap distance between the interface and the spherical colloid.
 }
\label{fig1_drag}
\end{figure}

\section{Theoretical framework}
\noindent \textbf{Brownian motion near a flat boundary; hydrodynamic friction.}
When a spherical particle of radius $a$ diffuses in a  homogeneous three-dimensional fluid of viscosity $\eta$, its diffusion coefficient is given by the Stokes-Einstein relation, % $D_0=K_B T/6\pi \eta a$
 $D_0=k_B T/\xi_0$.  
 Here,  $\xi_0=6\pi a \eta$ is the translational drag coefficient, $k_B$ is the Boltzmann constant and $T$ is  the temperature.  However, if the particle is close to an interface, the particle diffusivity is modified depending on the boundary conditions at the interface \citep{prieve1999measurement,lauga2005brownian}.
An exact and asymptotic solution of Stokes equation for a spherical particle moving close and parallel to an interface with no-slip and full-slip boundary conditions is known in the literature  \citep{goldman1967slow, o1964slow,lee1979motion, lee1980motion}. 
Experiments show good agreement with hydrodynamic predictions \citep{wang2009hydrodynamic,  lisicki2014translational, holmqvist2007colloidal,
walz1995study}.  In this work, we use approximate expressions  introduced in Ref. \citep{nguyen2004exact} for the drag coefficient of a colloidal sphere near a planar interface.  
 For a solid interface with no-slip boundary condition, 
 the hydrodynamic viscous drag in the parallel direction can be approximately described by \citep{goldman1967slow, nguyen2004exact},
\begin{equation}  \label{eq_xi_solid}
\xi_\parallel = \xi_0\lbrace 1+0.498\lbrace \log[1.207(a/h)^{0.986} +1]\rbrace^{1.027}\rbrace^{0.979},
\end{equation} 
where $h$ is the gap distance between the particle and the interface.
For full-slip boundary conditions the parallel drag can be written as \citep{nguyen2004exact},
\begin{equation} \label{eq_xi_liquid}
\xi_\parallel = \xi_0\frac{1.106+h/a}{1.501+h/a},
\end{equation}
 As illustrated in Fig.~\ref{fig1_drag}, the drag coefficients in both cases  approaches the bulk value for gap distances $h \gg a$, i.e., much larger than the particle radius,  and they behave differently for $h \ll a$, i.e., for gap distances much smaller compared to the particle radius. In the case of a planar interface with no-slip boundary conditions  or rigid wall, the drag coefficient  increases considerably when the particle gets closer to the interface. For a full-slip interface, however, the drag is always smaller than its bulk value and it decreases gradually  while the particle approaches to the interface.  
 
\begin{figure}
\includegraphics[width=1\columnwidth]{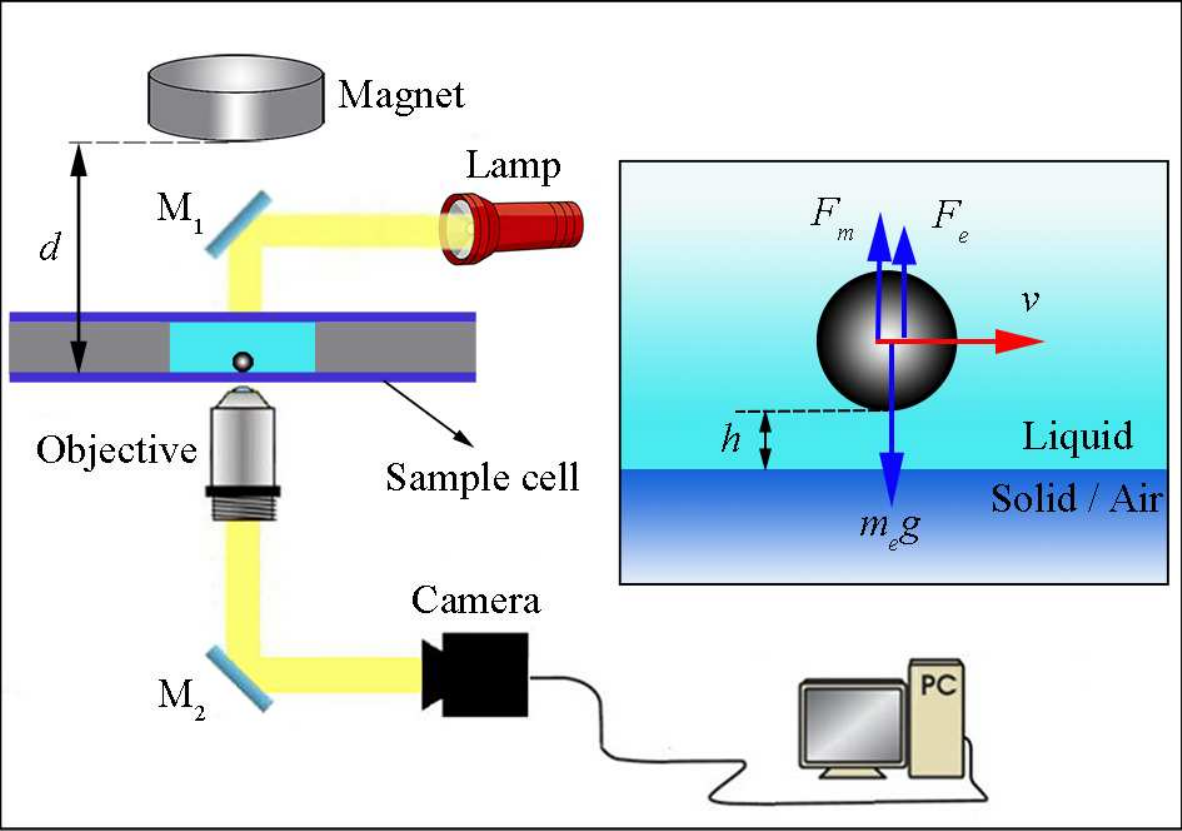}
\caption{ Schematic of the experimental setup for investigating the effects of the particle weight on its dynamics near an interface.  Inset: Colloidal magnetic particle moving with velocity $v$ at an interface forming between a liquid and either a solid or air. An upward magnetic force, $F_m$, is used to reduce the effective weight of the particle. 
Here, $F_e$ is the electrostatic repulsive force and $m_eg$ is the gravitational force minus buoyancy (see the text for the definition).}
\label{setup}
\end{figure}

The diffusion coefficient can be determined by a particle trajectory through the mean square displacement (MSD), which is defined as $\mathrm{MSD}(t) \equiv \langle \left[ {\mathbf r}(t) - {\mathbf r}(0) \right]^2 \rangle$, where $ {\mathbf r}$ is the particle position, $t$ is the time interval between initial and final position and $\langle \cdots \rangle$ denotes the ensemble average  over the equilibrium distribution \cite{bian2016111}.
 For two dimensional Brownian particles, it has been shown that $\mathrm{MSD}(t) = 4 D t $. The diffusion coefficient for parallel motion is found using the Einstein relation      $D_\parallel=k_B T/\xi_\parallel$. In the following we denote the  diffusion coefficient in parallel direction simply by $D$.  
\\ \\
\noindent \textbf{Magnetized particle in an external field.}
We consider a small spherical paramagnetic particle of radius $a$, located at a distance $h$ from a permanent magnet, as illustrated in Fig.~\ref{setup}. The particle is magnetized under the external magnetic field $\mathbf{B}$, with magnetization $\mathbf{M}$ \citep{jackson2007classical},
\begin{equation} 
\mathbf{M}=\frac{3}{\mu_0}\frac{\mu-\mu_0}{\mu +2\mu_0} \mathbf{B}
\label{magnetization}
\end{equation} 
where $\mu$ is the permeability of the substance forming the particle and $\mu_0$ is the vacuum permeability. 
Owing to its small size, the particle can be considered as a magnetic dipole with dipole moment of  $\mathbf{m}=(4\pi a^3/3) \mathbf{M}$. The magnetic force can then  be given by \citep{jackson2007classical},
\begin{equation} 
\mathbf{F}_m=\nabla (\mathbf{m}\cdot \mathbf{B})= \frac{4\pi a^3}{\mu_0}\frac{\mu-\mu_0}{\mu +2\mu_0} \nabla (\mathbf{B}\cdot  \mathbf{B})
\label{Force_magnet}
\end{equation}
which shows that the magnetic force acting on a dipole is proportional to $\mathbf{B}\cdot \nabla \mathbf{B}$.  

In our experiments, to control the effective weight of the colloidal particle, a permanent magnet was used to apply an upward force  $F_m$ (see the inset of  Fig.~\ref{setup}). This  is in fact equivalent to reducing the gravitational forces, directly allowing us to study the effects of the effective weight on the motion without changing any geometrical parameters in the system. In this work, the magnetic force is less than the gravitational force such that the particle  always tends to settle down towards the interface. The effective weight is,
\begin{equation}
F_{w,~eff} =m_e g-F_m
\label{Force_weight}
\end{equation}
where $m_e g$ is the effective gravitational force (gravity minus buoyancy, or  $F_g-F_b$), $m_e = \frac{4}{3} \pi a^3 (\rho_p-\rho_f)$ is the effective mass of the particle with density $\rho_p$ in a fluid of density $\rho_f$, and $g$ is the  gravitational acceleration.   
The electrostatic repulsion force between the particle and the interface prevents the particle to penetrate to the interface.
When the particle is far enough from the interface, the electrostatic repulsion is much smaller than the effective weight and the total force is directed downwards. When, instead, the particle is close enough to the interface, the electrostatic repulsion becomes larger than the effective weight and the total force is directed upwards. This allows the particle to sediment around an equilibrium gap distance, which depend on the relative strength of the effective weight and the electrostatic repulsion. After the particle has reached the equilibrium position, the force due to the thermal fluctuations makes the particle to fluctuate around that position.
\\ \\
\noindent \textbf{Brownian dynamics simulations.}
We complemented and validated our experimental observation by comparing them with numerical  simulations.
We implemented a Brownian dynamics simulation using an overdamped Langevin equation, 
\begin{equation}
{\mathbf v} = \frac{{\mathbf F}}{\gamma} + \sqrt{2 D}\,\mathbf {\dot{W}} + \mathbf{v}_{\mathrm{sd}}
\label{eq_Langevin}
\end{equation}
using the approach of Ref.~\cite{volpe2013simulation,callegari2019numerical}.
Here, ${\mathbf v}$ is the velocity of the particle,  $\gamma$ is the viscous friction coefficient, ${\mathbf F}$ is the net external force acting on the particle, $\mathbf {W}$ is the white noise with zero mean and variance of one,  and $\mathbf{v}_{\mathrm{sd}}$ is the spurious drift \cite{volpe2016effective}.
We consider the following forces acting on the particle: weight, buoyancy, magnetic force, and electrostatic colloidal force, i.e., ${\mathbf F} ={\mathbf F}_g +{\mathbf F}_b  + {\mathbf F}_m +  {\mathbf F}_e$.  The electrostatic colloidal force, ${\mathbf F}_e$, can be determined by  the Derjaguin-Landau-Verwey-Overbeek (DLVO) theory of colloidal stability \cite{israelachvili1992intermolecular, paladugu2016nonadditivity}:
\begin{equation}
F_{\rm e}(h) = \frac{k_{\rm B}T}{l_{\rm D}} \text{exp}\left( -\frac{h-l_{\rm e}}{l_{\rm D}}\right) 
\label{Force_electric}
\end{equation}
where $l_{\rm D}$ is the Debye length of the solution, and $l_e$ is a measure of the electrostatic interactions between the particle and the interface. 
\\ \\
\noindent \textbf{\bf Equilibrium distribution.} The probability of finding the particle at any  gap distance depends on the potential energy at that gap distance, which is determined quantitatively using Boltzmann's equation \citep{alexander1987hydrodynamic}:
\begin{equation}\label{Boltzmann}
P(h)=A\exp\left(- \frac{U(h)}{k_{\rm B} T}  \right) 
\end{equation}
where $P(h)dh$ is the probability of finding the particle at a location between $h$ and $h+dh$, $U(h)$ is the potential energy of the particle at gap distance $h$, and  $A = 1/Z$, where $Z$ is the partition function.  The average gap distance $\langle h \rangle$ is defined as
\begin{equation}
\langle h \rangle = \int h P(h) dh .
\end{equation}
When the potential has a narrow energy minimum, the $\langle h \rangle$ is {\em de facto} equal to $h_{\mathrm{eq}}$, i.e., to the gap distance where $U(h)$ has its minimum and the forces are in equilibrium.

\section{Experimental details}
The sample cell for liquid-solid interface was formed in between two glass slides separated by a $250~\mu \rm{m}$ parafilm spacer. 
For liquid-air experiments, a small vertical tube with a conical internal shape was used in which a droplet of the sample solution was placed at its bottom with an inner diameter of $4\,{\rm mm}$ \citep{aporvari2020anisotropic}. The paramagnetic particles in the solution (Microparticles PS-MAG-S2180, diameter $3.9\,{\rm \upmu m}$, density $1.62 \,{\rm g/cm^3}$, mass susceptibility $2.5\times10^{-4}\, {\rm m^3/kg}$) sedimented to interfaces because of gravity. 
A disk-shaped permanent magnet of 45 mm diameter and 15 mm height was used to apply a vertical magnetic force. 
For a permanent magnet in the shape of right circular cylinder, the magnetic field can be analytically determined at all points on the axis of the cylinder using the magnet characteristics \citep{jackson2007classical}. In our experiments, to accurately calculate both the magnetic field and its gradient at any point on the magnet axis, we used the theoretical magnetic field for a cylindrical magnet that was fitted  to measured fields at several points on the axis of the magnet.

To control the viscosity of the fluid, we used  methyl cellulose (M-280, 4000 cp, Sigma Aldrich, molar mass $\approx 88000$) and glycerol (99 \%, abcr GmbH \& Co. KG, density $1.26 \,{\rm g/cm^3}$, molar mass 92.11). A mixed solution of either methyl cellulose (MC) or glycerol with pure water was used to achieve different viscosities by varying the concentrations.
Low concentration MC solutions were  prepared by making first a 1 \% (wt/vol) solution in pure water. Solutions with different viscosities were then prepared by diluting the 1 \% solution with water. In this way, we prepared a series of MC solutions in semidilute regime with concentrations between $c^\ast$ and $3c^\ast$, where $c^\ast = 0.064~\%$ (wt/vol) is the overlap concentration\footnote{ 
 To estimate the overlap concentration, we first note that in diluted samples MC molecules form coil-like chains in water where their \emph{gyration} radius, $R_\mathrm{g}$, can be calculated using their molar mass \citep{funami2007thermal}. $R_\mathrm{g} \approx 38$ nm for the molar mass that we used in our experiments. The overlap number density, $n^\ast$, at which the polymer coils overlap is $1/v_p$, where $v_\mathrm{p}=(4\pi/3)R_\mathrm{g}^3$ is the coil volume \citep{lekkerkerker2011depletion}. The overlap concentration in terms of $\mathrm{ kg/m^3 }$ can then be written as
\begin{equation}
c^\ast=\frac{3M_p}{4\pi R_g^3 N_{av}}  
\end{equation} 
where $M_\mathrm{p}$ is the polymer's molar mass and $N_\mathrm{av}$ is Avogadro's number. Using the above average gyration radius and molar mass, the overlap concentration is $0.064$ \% (wt/vol).
}.
The viscosity of each solution was measured by means of a rheometer (MCR 301, Anton Paar) at lab temperature (see Appendix). 
 To calculate the buoyancy forces in glycerol samples, we measured the density of each sample with different concentration separately. 
 The difference between the density of pure water and MC or salt solutions in our experiments were negligible due to extremely low concentrations of the additives. However, in glycerol samples where we worked at higher concentrations, we directly measured the density of each sample with different concentrations (see Appendix).  

The samples were imaged using a custom-built microscope consisting of a $20\times$ objective lens (${\rm NA}=0.50$, ${\rm WD}=0.17\,{\rm mm}$) and a digital camera (Thorlabs DCC1645C). 
We recorded videos of isolated particles at 120 frames per second for about 80 seconds. Recording at  a relatively high frame rate allowed us to improve the precision of the measurements by reducing the effect of unwanted drift at short-time regime.  
The videos  were analyzed with a custom digital video microscopy code in MATLAB \cite{crocker1996methods}.

\section{Results and discussion}
We investigated experimentally the case of a super-paramagnetic colloidal particle suspended in (a) water and (b) semi-diluted MC-water solution for both the cases of  (i) liquid-solid and (ii) liquid-air interface as a function of the effective weight, i.e., for different values of the magnetic field.
For the case of a particle suspended in water, both the liquid-solid and the liquid-air interface give the expected results, in agreement with the prediction of the hydrodynamic model.

For the semi-diluted MC-water solution, we find that the liquid-solid interface gives results that are in agreement with the hydrodynamic model for D corresponding to no-slip boundary condition. On the other hand, we find that, in the case of a colloid suspended in a semi-diluted solution of MC-water, its extrapolated diffusion coefficient for a liquid-air interface is not consistent with the prediction based on a full-slip boundary condition at the interface, rather it is more compatible with the trend in the presence of no-slip boundary condition.

\begin{figure}
\includegraphics[width=0.8\columnwidth]{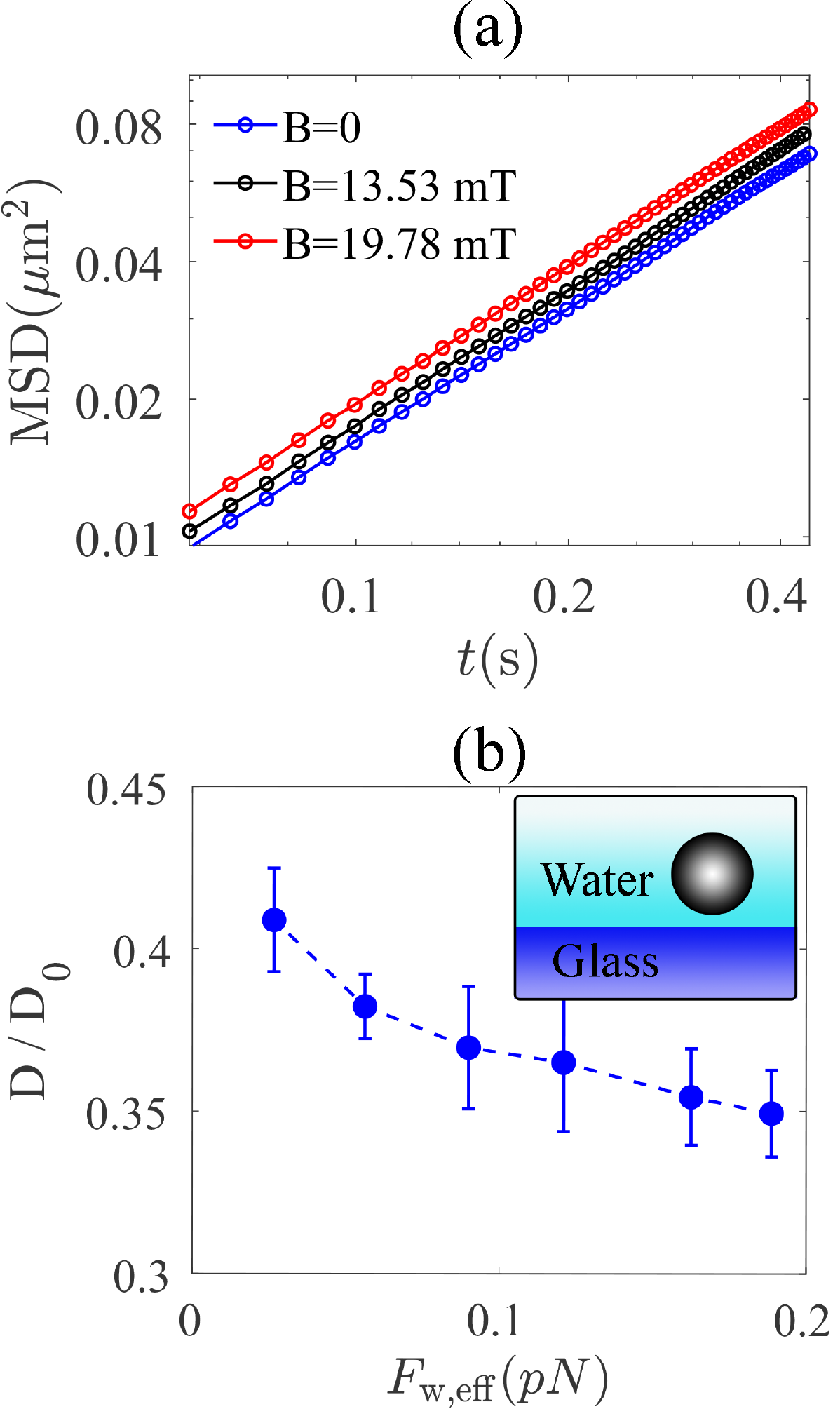}
\caption{(a) Typical MSDs of a magnetic particle for three different normal magnetic fields ($D =$ 0.040, 0.044, and 0.049 $\mu m^2/ s$ for blue, black and red lines, respectively) (b) Diffusion coefficients as a function of effective weight ($D_0=k_B T/\xi_0$, where $\xi_0$ is the Stokes friction coefficient). The error bars show one standard deviation across different measurements. The inset shows a sketch of a colloidal particle at a water-glass interface.  }
\label{fig3}
\end{figure}

To investigate if this trend is merely due  to the relatively high viscosity of the MC-water solution employed in the experiments we compared the diffusion of a particle in MC-water solution at liquid-air interface with one immersed in a water-glycerol solution of comparable viscosity. Also, to investigate  whether it  is instead related to a molarity effect, given that MC-water solutions are composed of small coil-like MC-polymer units of a few tens of nanometers suspended in water, we compared also with the case of a particle suspended in NaCl-water solution at a liquid-air interface.
\\ \\
\noindent \textbf{Water solution, liquid-solid interface.}
We first consider a single magnetic particle that sedimented on a water-glass interface as shown in Fig.~\ref{setup}. 
In Fig.~\ref{fig3}a, the MSDs of a magnetic particle at water-glass interface is shown for three different perpendicular magnetic fields. As shown in Fig.~\ref{fig3}a, the MSD curves in logarithmic scale shift upward for increasing applied magnetic field indicating an increase in diffusivity. 
In other words, reducing the effective weight by increasing the upward magnetic forces leads to an increase in the diffusivity of the particle on the water-glass interface.

 To better highlight the effect of the particle weight on its diffusivity at liquid-solid interface, we measured diffusion coefficients, $D$, of a single particle as a function of the effective weight. In all experiments the magnetic forces were chosen to be smaller than the gravitational forces such that the particle always stayed at the interface. 
In all experiments MSDs were fitted into a theoretical model $4Dt$. The results are shown in Fig.~\ref{fig3}b, where a nonlinear dependence on the  effective weight can be seen. 
This trend is in agreement with the variation of the hydrodynamic drags as a function of the gap distance $h$, between the particle and interface (see the comparison with the Brownian dynamic simulation). 
Note that, even for very small effective weight that are considered here, the particle is still in close contact with the interface and the gap distance, $h$, is in the range of few hundred nanometers,  as confirmed below using Brownian dynamic simulations.

%\\ \\

\begin{figure}
\includegraphics[width=0.8\columnwidth]{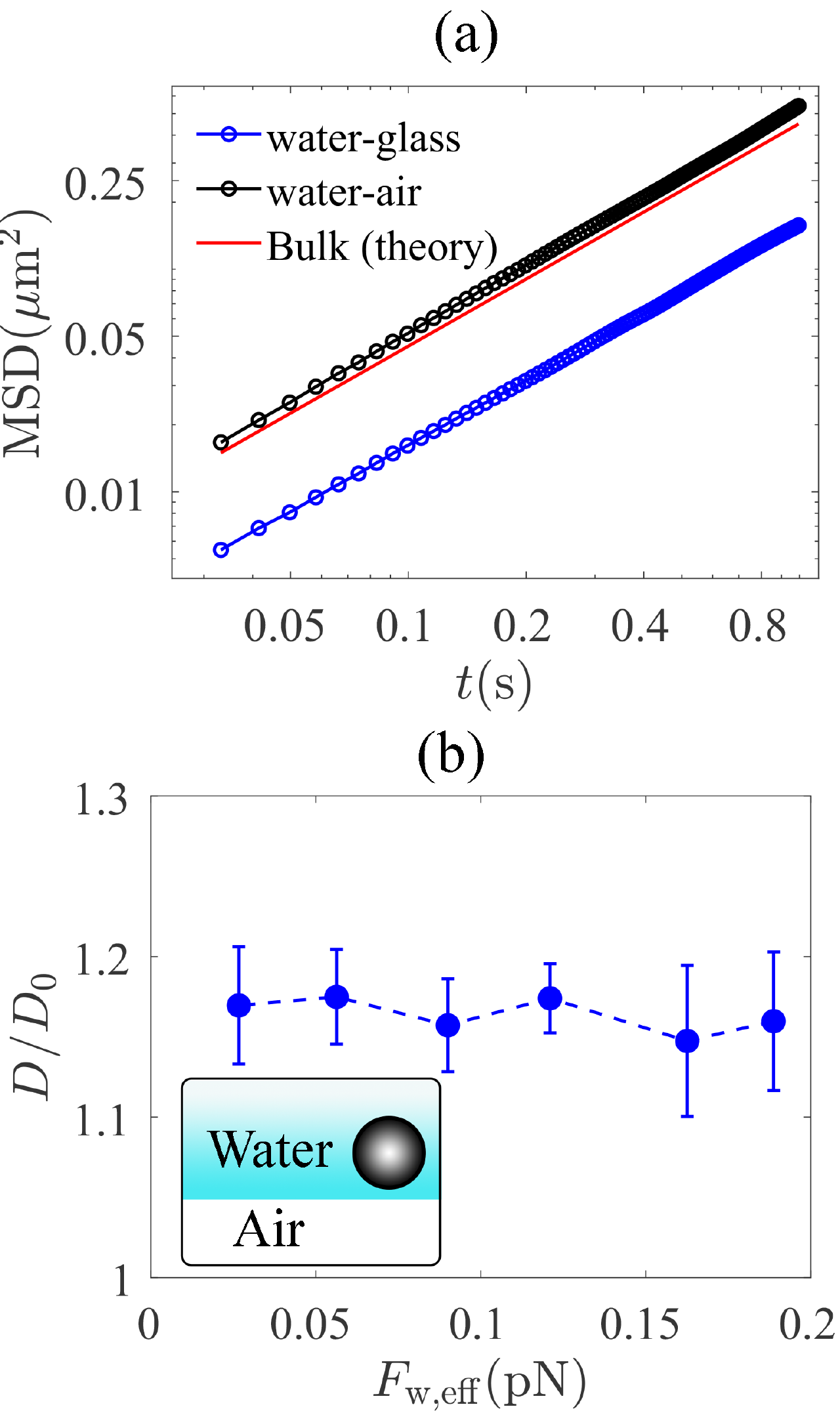}
\caption{(a) Typical MSDs for a magnetic particle at water-air (black circles, $D=0.130 ~ \mu m^2/ s$) and water-glass (blue circles, $D=0.040~ \mu m^2/ s$) interfaces for $\mathbf{B}=0$. The red line shows the corresponding theoretical  MSD (Stokes-Einstein relation) in bulk water. (b)  Diffusion coefficients as a function of the effective weight.  The error bars show one standard deviation across different measurements. The inset shows a sketch of a colloidal particle at a water-air interface. } 
\label{fig4}
\end{figure}

\begin{figure*}
\includegraphics[width=2\columnwidth]{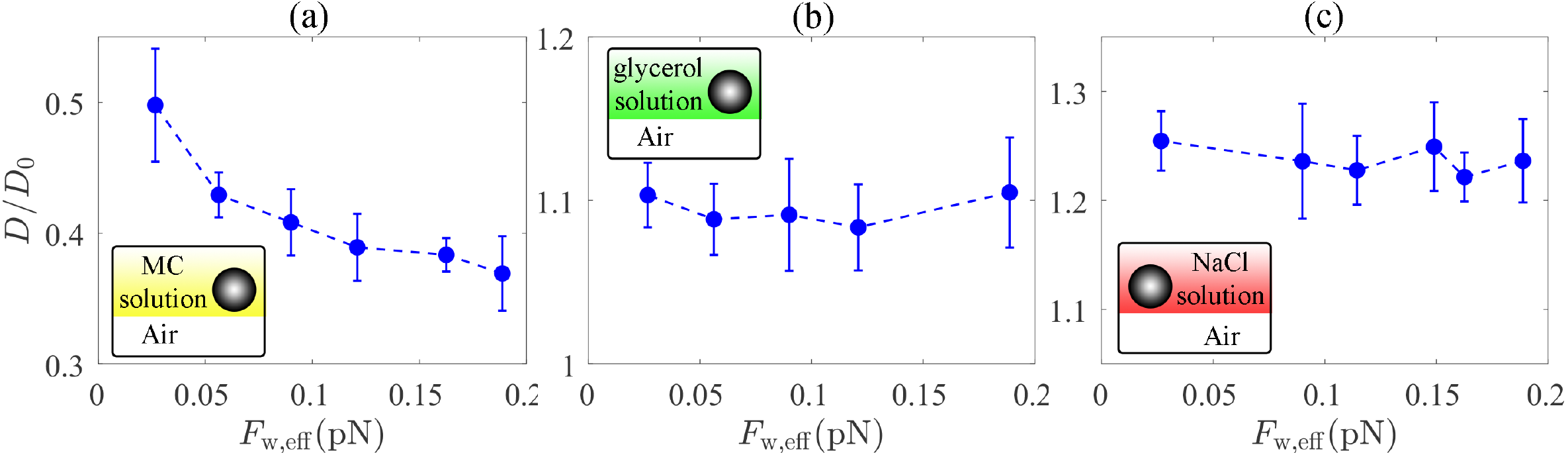}
\caption{ Diffusion coefficients at MC solution-air (a), glycerol solution-air  (b), and  sodium chloride  solution-air (c) interfaces. The concentration of the MC solution is  0.175\% (wt/vol)  with viscosity $\eta=3.79$ cp. The glycerol solution consists of 20 \% (vol/vol) glycerol  with viscosity  $\eta=2.13$ cp and the sodium chloride  solution consists of 20 mM NaCl with nearly the same viscosity as pure water.}
\label{fig5}
\end{figure*}
%\\ \\
\noindent \textbf{Water solution, liquid-air interface.}    
Next, we focus on   liquid-air interfaces.   
Due to the very large water-air tension at water-air interface, the deformation of the interface caused by the effective weight of the colloidal particle is negligible and hence the interface can be  considered to be flat \cite{mbamala2002effective}.  
Fig.~\ref{fig4}a shows representative MSDs of a magnetic particle in both water-air and water-glass interfaces with no applied magnetic field, both demonstrating  linear behavior. Here, for comparison, we also show the theoretical MSDs in the bulk of a water solution  based on  the Stokes-Einstein relation. While in the case of water-glass interface MSDs are considerably smaller than those in the bulk due to large interface frictional forces, for a water-air interface they are slightly bigger than the bulk, which is in agreement with Eq. \ref{eq_xi_liquid}  and the results in the literature \cite{wang2009hydrodynamic}. 
Also, notice that the MSD values at liquid-solid interface are about 3 times smaller  than the corresponding values in the bulk, which is due to the bigger parallel drag coefficients at sufficiently small gaps as can be confirmed in Fig.~\ref{fig1_drag}.
In liquid-air case, as can be seen in Fig.~\ref{fig4}b, the diffusion coefficients do not change with effective weight in a water-air interface, at least within the range of our experimental errors.  This behavior can be explained by considering Fig.~\ref{fig1_drag}:  While the viscous friction coefficient strongly changes by the gap distance $h$ at liquid-solid interface, it does not change considerably at liquid-air interface specifically at  gap distances $h < 0.2a $ where $a$ is the radius of the colloid.  %at very small gap distances.
\\ \\
\noindent \textbf{MC-Water solution, liquid-air interface.} 
We now consider  a colloidal particle  in an aqueous MC solution at semi-dilute regime. 
MC is the simplest water soluble cellulose derivative.   Aqueous solutions of MC are Newtonian at room temperature  especially for low concentrations that we used in our experiments \citep{nasatto2014interfacial, funami2007thermal}. Also, within experimental uncertainties, we did not observe any noticeable nonlinear behavior in the MSDs of suspended particle in our MC solutions, i.e. $\mathrm{MSD} \propto D t^\alpha$, with $\alpha =1$. 
In Fig.~\ref{fig5}a, MC solution with polymer concentration of $c =0.175 \%$ (wt/vol) ($c= 2.7~c^\ast $) is used to form a liquid-air interface.  
Interestingly, the diffusivity of a colloidal particle at the interface between semidilute MC solution and air is similar to those at  liquid-solid interfaces (Fig.~\ref{fig3}):  the diffusion coefficient reduces with increasing effective weight.
This could be due to the complex interaction between the colloid and  the crowded surrounding environment with MC molecules.
In fact, MC as an amphiphilic polymer adsorb at the water-air interface, where they can form polymer monolayers \citep{nasatto2014interfacial, nasatto2015methylcellulose, sarkar1984structural}. Also, MC can adsorb on the particle surface as well \citep{nasatto2014interfacial}. The adsorbed layers provide corrugated interfaces that  enhance the Van der Waals interactions \citep{nasatto2015methylcellulose},  which could be responsible for the effect that we observed here through increasing the resistance forces between the particle and the interface.  
Hence, we hypothesize that the adsorb polymer monolayer at water-air interface changes the boundary conditions at the water-air interface from  perfect slip into a partial slip boundary condition. 
\\ \\
\noindent \textbf{Glycerol-Water solution, liquid-air interface.}  
To see that the unexpected trend of $D$ in the MC solution-air is not due to the viscosity of the fluid, we performed experiments using solutions of glycerol, which has small polyol molecules ($\sim 10 $ \r{A}) as opposed to the large molecules of MC. The result is shown in Fig.~\ref{fig5}b  for a 20 \% (vol/vol) glycerol solution in pure water (viscosity $\eta=2.13$ cp). As shown in Fig.~\ref{fig5}b, here the diffusion coefficient does not change with the effective weight in the range of the experimental errors. According to these results,  
the glycerol solution-air interface has the same interface boundary conditions seen for the pure water-air interface.
\begin{figure*}
\includegraphics[width=2\columnwidth]{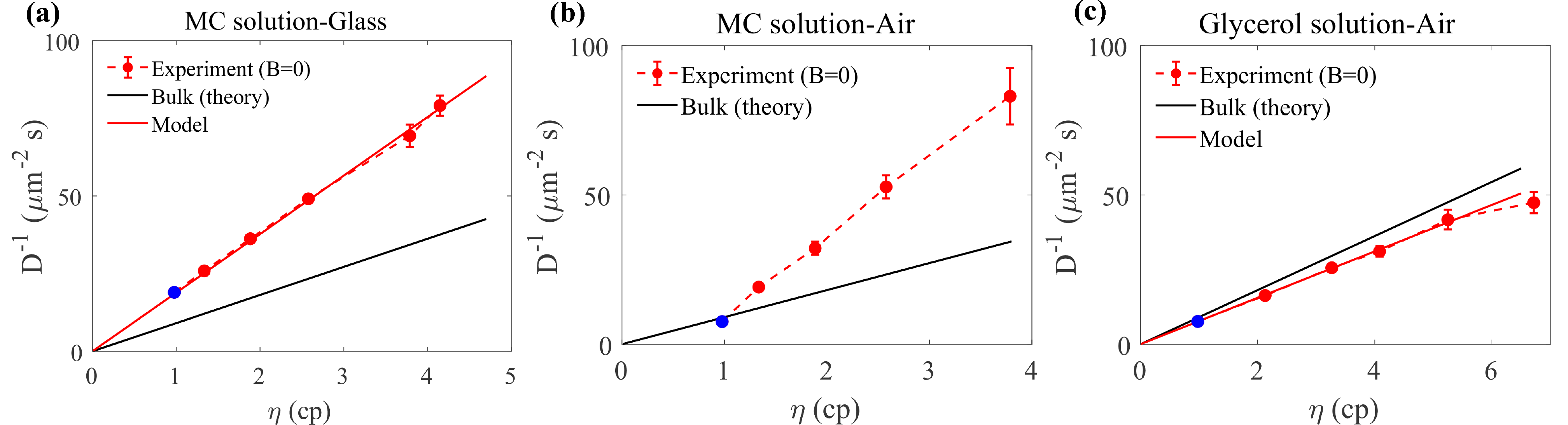}
\caption{ Inverse of the diffusion coefficients of a magnetic particle as a function of viscosity of  the solutions with different concentrations in (a) MC solution-glass, (b) MC solution-air, and (c) glycerol solution-air interfaces. The black line in each panel shows the theoretical values in a bulk fluid from the Stokes-Einstein relation.
 Note that the density of glycerol solutions is variable depending on its concentration, which leads to variable effective weight in this case. However, the difference of the density of MC solutions with pure water is negligible due to their low concentrations in our experiments.  
 The highest concentration of MC in the samples is 0.2 \% (wt/vol)  while it is 45 \% (vol/vol) for glycerol. Blue dot in panel (a) shows the dynamics in pure water-glass interface and in panels (b) and (c) show the dynamics in pure water-air interface.  
   }
\label{fig6}
\end{figure*}
\\ \\
\noindent \textbf{NaCl-Water solution, liquid-air interface.}  
 Next, to investigate the effect of non-hydrodynamic effects of the particle dynamics, we consider the effect of the molarity of the fluid on the particle diffusivity. By adding sodium chloride (NaCl), we can change the electrostatic properties of the interface \citep{butt2010surface}.  Fig.~\ref{fig5}c shows diffusivity at interface between 20 mM NaCl and air:  a high  molarity sample of NaCl does not substantially affect the diffusivity of a magnetic colloid at interface compared to the pure water case (see Fig.~\ref{fig4}b). In  this case the average diffusion coefficient is somewhat bigger than the corresponding value in the pure water case, which may be related to smaller equilibrium distance between the particle and the interface at high molarities \cite{villa2020multistable} (which is what one would predict using electrostatic consideration and the fact that the Debye screening length in an electrolyte solution is smaller than that in pure water). Because the drag forces are smaller at smaller gap distances between the particle and full-slip interface \cite{wang2009hydrodynamic}, here the diffusivity of the particle is higher than that at the pure water-air interface. 
\\ \\
\noindent \textbf{Dependence on the viscosity at $B=0$.}
 It is customary to investigate the dependence of the inverse of the diffusion coefficient ($D^{-1}$) on the fluid viscosity.
  In fact,  $D_\parallel^{-1}= (6 \pi a/K_\mathrm{B}T) \eta f_\parallel$ with $f_{\parallel}=\xi_{\parallel} /\xi_0$,  hence $D_\parallel^{-1}$ is proportional to the viscosity $\eta$ and the  normalized drag coefficient, $f_\parallel$. 
Fig.~\ref{fig6}a and b show  the inverse of the diffusion coefficients %for three different normal forces 
as a function of the viscosity of MC solutions with different concentrations, where $\eta=0.98$ cP corresponds to the pure water case at $20^\circ$ C (blue circle in the figure). Here, we do not consider the effect of the magnetic field on the dynamics of the particle, i.e., $B=0$.  In  MC-glass interface (see Fig.~\ref{fig6}a), the diffusion coefficients considerably deviate from their bulk values in all viscosities due to the large frictional forces at the solid interface. 
 As we discussed previously, $f_\parallel$ is strongly dependent on the interface boundary condition when the particle is in close contact with the interface. At liquid-solid interface, we expect that  $f_\parallel$ does not change substantially with the viscosity of the fluid and  $D^{-1}$ is a linear function of $\eta$ . This is shown in Fig.~\ref{fig6}a where the experimental data in the absence of the upward magnetic field (i.e., corresponding to the highest effective weight  $F_{w,eff}=0.189$ pN), are fitted to a linear function (red solid line). By comparing the slope between this line and the bulk theoretical line and using Eq. \ref{eq_xi_solid}, our experiments show that the gap distance between  the particle and glass surface without the upward magnetic force  is about 390 nm, a reasonable value compared to the reported values in the literature \citep{villa2020multistable}.

The diffusivity as a function of viscosity  at MC-air interface in Fig.~\ref{fig6}b, shows a more complex behavior. At low viscosity, $D_\parallel$ is not too far from the bulk values. At large viscosities however, it demonstrates a large deviation from the bulk values. This is consistent with the hypothesis that  the adsorbed monolayer at interface changes the boundary condition of water into partial-slip boundary condition. At higher MC concentration we expect that the boundary condition approaches toward the no-slip boundary condition for solid surfaces. In this case, $f_\parallel$ is no longer a constant and varies by the viscosity $\eta$,  which  makes the diffusion coefficient a nonlinear function of viscosity 
 (supposing $D^{-1}$ starts from zero at $\eta =0$ ). 

Next, we consider glycerol (as a monomer) to control the viscosity of the solutions instead of a polymer chains like MC.
 The result is shown in Fig.~\ref{fig6}c. As can be seen, here $D_\parallel^{-1}$ is less than the corresponding values in the Stokes-Einstein relation for bulk fluid,  which is the typical behavior at interfaces with slip boundary conditions (see Fig.~\ref{setup}). Remarkably, it shows a linear dependency on the viscosity,
indicating that the small glycerol molecules do not considerably change the hydrodynamic boundary conditions seen for water-air interface.  This result is also in agreement with the results in the Fig.~\ref{fig5}b where the diffusivity at the glycerol-air interface shows the same behavior as the pure water case.
Note that in Fig.~\ref{fig6}c the density of the samples is a function of glycerol concentration leading to variable effective weights in different samples, which affects the average distance between the particle and the interface at different viscosities. Therefore,  we cannot infer the gap distance from the plot.  Nonetheless, also in this case  $f_\parallel$ does not change considerably with the gap distance at interfaces with full-slip boundary conditions leading to a linear relation between $D^{-1}$ and $\eta$ as shown in Fig.~\ref{fig6}c. 

\begin{figure*}
\includegraphics[width=1.8\columnwidth]{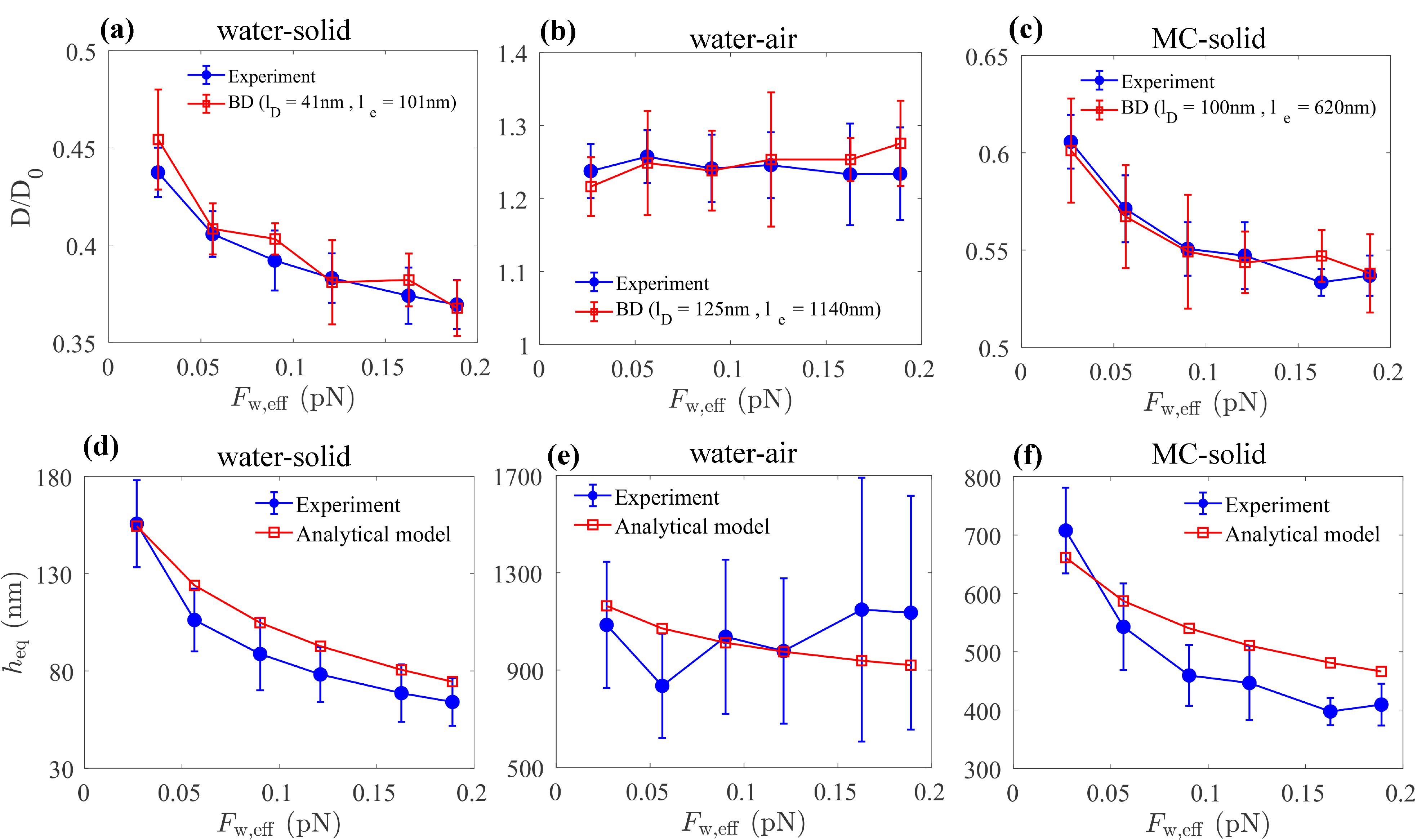}
\caption{Brownian dynamics (BD) simulation for (a) water-solid (b) water-air and (c) MC-solid interfaces using DLVO electrostatic force for different electrostatic parameters from Eq. \ref{Force_electric}. At liquid-solid interfaces we chose no-slip boundary conditions while at liquid-air interfaces we used full-slip boundary conditions. Blue circles show the corresponding experimental results. In the experiments, the solid surface was glass and the viscosity of MC solution was  $\eta =3.79$ cp.       
(d)-(e) show the corresponding equilibrium gap distance, $h_\mathrm{eq}$, as a function of the effective weight; red squares show the gap distances that are calculated using an analytical model (see the text) and blue circles show the experimental gap distances that were calculated using  Eqs. \ref{eq_xi_solid} and  \ref{eq_xi_liquid}. 
}
\label{fig7}
\end{figure*}

%\\ \\
\noindent \textbf{Simulation and theoretical model.}
Finally, to rationalize our experimental findings, we implemented three dimensional Brownian dynamics simulations by solving Eq. \ref{eq_Langevin} using Eqs. \ref{Force_magnet},  \ref{Force_weight}, and \ref{Force_electric}.  
In the simulations, the electrostatic parameters from Eq. \ref{Force_electric}, i.e. $l_\mathrm{D}$ and $l_\mathrm{e}$,   were set as fitting parameters to reproduce the experimental conditions. All other parameters  were chosen to have the same values as in the experiments.  
To compare with the results of the experiments, 10 different trajectories were used in each simulation. 
Depending on the trend shown experimentally by the $D_\parallel$ as a function of $F_{w,eff}$, we used drag coefficients from Eq. \ref{eq_xi_solid} or \ref{eq_xi_liquid}    for  liquid-solid or liquid-air interfaces, respectively.
Fig.~\ref{fig7}a shows the simulation results at water-solid interface assuming no-slip boundary condition. As can be seen there is good agreement between the simulation and the experimental results for the chosen electrostatic parameters. 
The water-air case is shown in Fig.~\ref{fig7}b where a full-slip boundary condition is used at the interface. Here,  similarly to the experiments the diffusion coefficients do not change considerably with the effective weight.

Also, according to the simulations, the Debye lengths are $l_\mathrm{D} \approx 41 \pm 5\, {\rm nm}$, and $l_\mathrm{D} \approx 125 \pm 10\, {\rm nm}$  for water-solid and water-air interfaces, respectively.
 $l_\mathrm{e}$ is even more sensitive to the interface properties as expected. While $l_\mathrm{e}\approx 101 \pm 10\, {\rm nm}$ at water-glass interface, it is considerably bigger at water-air interface ($l_\mathrm{e}\approx 1140 \pm 30\, {\rm nm}$).  

 We also performed similar simulations for MC  solutions. Again, as can be seen in Fig.~\ref{fig7}c there is  a good agreement between the simulations and the experiments for MC solution-solid interface. We used no-slip boundary condition in the simulations at liquid-solid interfaces. For the MC solution-air case (not shown here), our simulations that were performed assuming full-slip boundary conditions are not consistent with the experimental results, confirming that the presence of MC polymers chains changes the boundary conditions to a partial slip. 

The equilibrium gap distance $h_\mathrm{eq}$, can be calculated analytically from the %potential energy $U(h)$
 net force on the particle using estimated electrostatics parameters $l_\mathrm{D}$ and $l_\mathrm{e}$ that were calculated in Fig.~\ref{fig7}a-b. 
Fig.~\ref{fig7}d-e show the corresponding equilibrium gap distance,   as a function of effective weight at different interfaces, which again confirms that the gap distance is a function of effective weight. 
For the MC-solid interface case, the simulations show that the gap distance is about 466 nm for $B=0$, which is in the same range that we calculated above based on the data in Fig.~\ref{fig6}a.
We also show the corresponding gap distances that were calculated  using  the experimental diffusion coefficients by implementing  Eqs. \ref{eq_xi_solid} and  \ref{eq_xi_liquid} (blue circles in Fig.~\ref{fig7} d-f). For liquid-solid interfaces, there is  a good agreement between the experimental and analytical values showing the consistency between the two models. In the case of liquid-air interface, the experimental values follow a constant trend because we are in a range of distances where the diffusion coefficient does not change appreciably by changes in $h$ on the order of hundreds of nm. (see Fig.~\ref{fig1_drag}).  The experimental and analytical values are in the same range.

\begin{figure*}
\includegraphics[width=1.8\columnwidth]{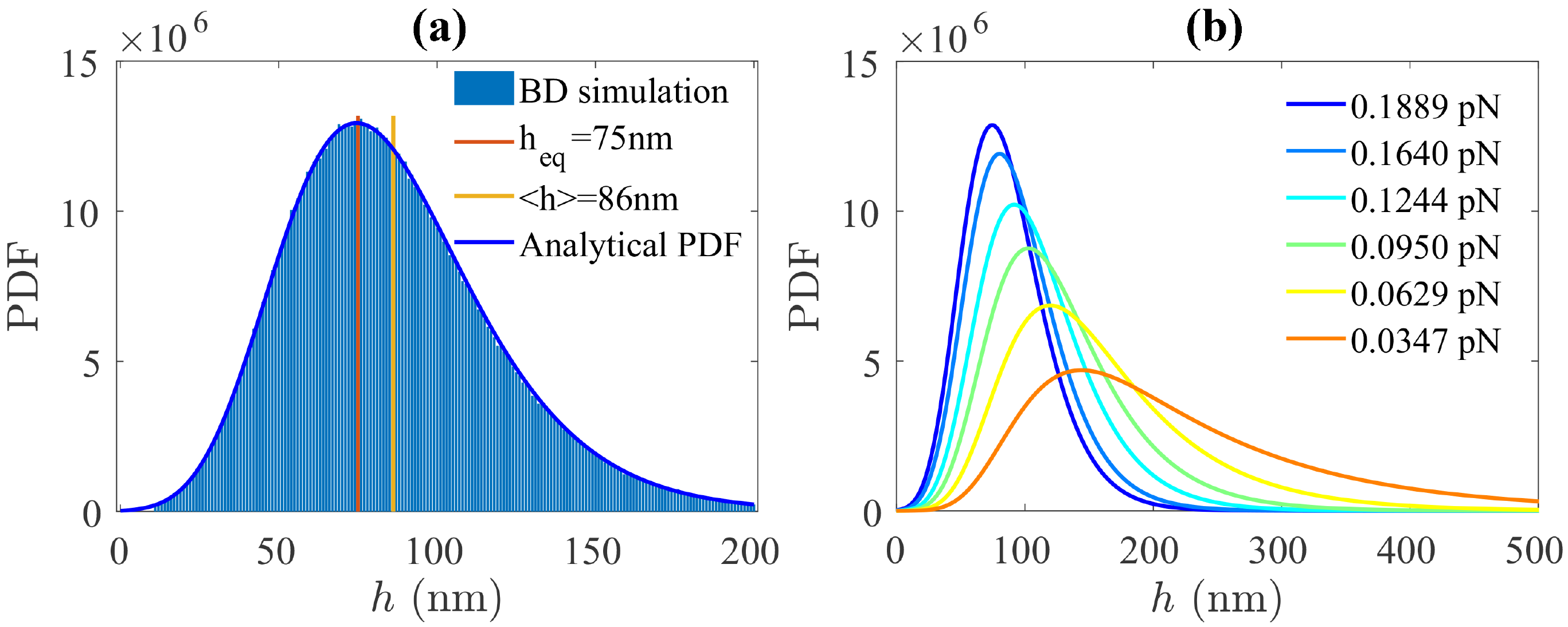}
\caption{(a) The PDF of the gap distances of the particle at water-solid interface for $B=0$ and electrostatic parameters calculated in Fig.\ref{fig7}a . Blue bars are extracted from 100 different trajectories  calculated using  Brownian dynamics (BD) simulation. Red and yellow vertical lines show the most probable gap distance and mean gap distance, respectively, calculated from BD simulations. Blue curve is the analytical PDF   from Eq. \eqref{Boltzmann}. (b) Analytical PDFs calculated for different effective weights. 
}
\label{fig_PDF}
\end{figure*}

We also use the parameters of the Brownian dynamics  simulations to calculate the probability density function (PDF) of the gap distances of the particle at water-solid interface. Fig.~\ref{fig_PDF}a shows PDF for the case of $B=0$ extracted from 100 different trajectories. In the simulations, the electrostatic force was calculated  based on the parameters extracted from Fig.~\ref{fig7}a. The blue curve shows the analytical PDF that was calculated from Eq. \eqref{Boltzmann} with the same electrostatic parameters. As  expected there is an excellent agreement between the BD simulations and the analytical distribution.   

Red vertical lines show the most probable gap distance, $h_\mathrm{eq} = 75$ nm, which is very close to the corresponding analytical value, i.e. 74.5 nm. The mean gap distance is  also shown (blue vertical line) in the figure, which differs from the most probable gap distance due to the asymmetric nature of the distribution.  Fig.~\ref{fig_PDF}b shows the analytical PDFs calculated for different effective weights. As can be seen, for smaller effective weights (corresponding to stronger magnetic forces) the distribution is more asymmetric with bigger equilibrium gap distance. Thus the gap distance is bigger for smaller effective weights at water-solid interface.

\section{Conclusions}
Development of new techniques to investigate the dynamics of colloidal particles at interfaces provides a promising way to understand the interfacial interactions.  In this work, we studied the effects of the weight of a colloidal particle on its dynamics in close contact with an interface using an upward magnetic force. In our system, particles are confined  in the direction perpendicular to the interface due to the gravity with a small gap distance where the gravity and  interfacial forces are balanced.  We have shown how the interplay between hydrodynamic, weight and electrostatic interactions  affect the dynamics of a spherical colloid at different interfaces. Consistent with the previous results in the literature, at liquid-solid interfaces the diffusion  coefficients obtained from particle trajectories considerably change by the colloid's weight due to the strong dependency of the hydrodynamic drag to the gap distance between the particle and the interface. 
At liquid-air interfaces, while for pure water the diffusion coefficients change  negligibly by gap distance, we have shown that an adsorbed polymer monolayer at water-air  interface changes the boundary conditions from full-slip to a partial-slip, as exemplified by the case of a particle suspended in a semi-diluted MC  solution.  
In other words, at the interface between a semidilute MC solution and air, the colloidal dynamics is more similar to those at the liquid-solid interfaces i.e., the diffusion coefficients change considerably by the colloid's effective weight. 
Our numerical simulations  have shown a  good agreement with experiments.

In this work we only considered the magnetic forces such that the particles remain at the bottom interface in the sample chamber.  However, our technique allows both decreasing and increasing the effective weight.  
If we increase the magnetic force to sufficiently large values, the particle will detach from the bottom interface and will trap to the upper interface of the cell chamber. In this case, at strong magnetic fields, we  could  study the diffusivity of the particle  at effective weights  with absolute values even bigger than $m_eg$, which would allow to explore the gap distances much closer to the interfaces.   
 The simple technique we introduce here could be used further to study interfacial interactions in many physical systems.

\section*{Appendix}

%\noindent \textbf{A Determining the diffusion coefficients from experimental data}
\subsection*{Determining the diffusion coefficients from experimental data}
In the experiments we recorded videos of isolated particles at 120 frame per second for about 80 seconds. To avoid the problem of potential  mass changes between different particles, we performed the experiments with the same particle at different magnetic fields.  For each magnetic field, we repeated the experiments ten times. The MSDs were calculated from the trajectories and fitted to the theoretical MSDs at short time regime to minimize the effect of unwanted drifts in the experiments. The average diffusion coefficients were calculated and one standard deviation was considered as experimental errors in the figures.  

%\noindent \textbf{B  Density and viscosity of glycerol and MC solutions}
\subsection*{Density and viscosity of glycerol and MC solutions}
The viscosities of glycerol and MC solutions  with different concentrations, $c$, used in this work are shown in Table \ref{grycerol} and Table \ref{MC}, respectively. The viscosities were measured using an Anton Paar rheometer (MCR 301). The densities of glycerol solutions were also measured directly and reported in  Table \ref{grycerol}. The densities  of MC solutions were very close to the density of pure water as their  concentrations were very low.   
All data were  measured at $20^\circ$C.
\begin{table}[ht]
\caption{Densities and viscosities of glycerol solutions with different concentrations.}
\centering

\begin{tabular}{c | c c c c c} 
%\hline
$c$ (\% vol/vol)         & 20   & 30   & 35   & 40   & 45   \\
\hline
$\rho$  (Kg/m$^3$)       & 1054 & 1081 & 1098 & 1111 & 1128 \\
\hline
$\eta$  (cp)             & 2.13 & 3.27 & 4.08 & 5.25 & 6.72 \\
%Viscosity \footnote{Based on the data in \cite{cheng2008formula}} (cp) & 1.98 & 2.99 & 3.76 & 4.81 & 6.28 \\ 
%\hline
\end{tabular}
\label{grycerol}
\end{table}

%methylecellulose
\begin{table}[ht]
\caption{Viscosities of MC solutions with different concentrations.}
\centering

\begin{tabular}{c | c c c c c} 
%\hline
% $c$ (\% vol/vol)              & 0.025 & 0.075 & 0.125 & 0.175 & 0.200 & 0.250   \\
  $c$ (\% wt/vol)              & 0.025 & 0.075 & 0.125 & 0.175 & 0.200   \\
%\hline
%Density \footnote{Measured at $20^\circ$C } ($\rm{kg/m^3}$) & 1054 & 1081 & 1098 & 1111 & 1128 \\
\hline
%$\eta$  (cp)                         & 1.34 & 1.89 & 2.58 & 3.79 & 4.15 & 5.88 \\
$\eta$  (cp)                         & 1.34 & 1.89 & 2.58 & 3.79 & 4.15 \\
%Viscosity \footnote{Based on the data in \cite{cheng2008formula}} (cp) & 1.98 & 2.99 & 3.76 & 4.81 & 6.28 \\ 
%\hline
\end{tabular}
\label{MC}
\end{table}

\section*{Author Contributions}
All authors discussed and interpreted results and contributed to the preparation of the manuscript. M.S.A. is the main author of the manuscript. He devised and conducted the experiments, analyzed the experimental data, ran the numerical simulations for the comparison with the experiments, and wrote the paper. 
A.C. was involved in the delineation of the theoretical framework, provided the code for the comparison with the Brownian dynamics simulation.
 E.U.S. supervised the project, and contributed to the design of the magnetic setup and the magnetic analysis.

%We strongly encourage authors to include author contributions and recommend using \href{https://casrai.org/credit/}{CRediT} for standardised contribution descriptions. Please refer to our general \href{https://www.rsc.org/journals-books-databases/journal-authors-reviewers/author-responsibilities/}{author guidelines} for more information about authorship.

\section*{Conflicts of interest}
There are no conflicts to declare
%In accordance with our policy on \href{https://www.rsc.org/journals-books-databases/journal-authors-reviewers/author-responsibilities/#code-of-conduct}{Conflicts of interest} please ensure that a conflicts of interest statement is included in your manuscript here.  Please note that this statement is required for all submitted manuscripts.  If no conflicts exist, please state that ``There are no conflicts to declare''.

\section*{Acknowledgements}
Helpful discussions with Joakim Stenhammar and Giovanni Volpe are kindly acknowledged. This work was supported by the Scientific and Technological Research Council of Turkey
(TUBITAK 215E198).

%The Acknowledgements come at the end of an article after Conflicts of interest and before the Notes and references.

%%%END OF MAIN TEXT%%%


\balance


%%%REFERENCES%%%
%\bibliography{rsc} %You need to replace "rsc" on this line with the name of your .bib file
\bibliography{Bibliography}
\bibliographystyle{rsc} %the RSC's .bst file

\end{document}